\documentclass[prb,showpacs,showkeys,preprintnumbers,superscriptaddress,a4paper,twocolumn]{revtex4-1}
\usepackage[T1]{fontenc}
\usepackage[utf8]{inputenc}
\usepackage{graphicx}
\usepackage{amsmath,amssymb}
\usepackage{amstext}													
\usepackage[colorlinks,citecolor=blue,
    filecolor=blue,
    linkcolor=blue,
    urlcolor=blue]{hyperref}

\def\beq{\begin{equation}}
\def\eeq{\end{equation}}
\def\bea{\begin{eqnarray}}
\def\eea{\end{eqnarray}}
\def\ba{\begin{array}}
\def\ea{\end{array}}

\begin{document}
\title{Entanglement and Majorana edge states in the Kitaev model}
\author{Saptarshi Mandal}\email{saptarshi@iopb.res.in}\affiliation{Institute of Physics, Bhubaneswar-751005, Orissa, India}
\author{Moitri Maiti }\email{maiti@theor.jinr.ru}\affiliation{BLTP, JINR, Dubna, Moscow region, 141980, Russia}
\author{Vipin Kerala Varma }\email{vvarma@ictp.it}\affiliation{The Abdus Salam ICTP, Strada Costiera 11, 34151, Trieste, Italy}

\begin{abstract}
We investigate the von-Neumann entanglement entropy and Schmidt gap in the vortex-free ground state of the Kitaev model 
on the honeycomb lattice for square/rectangular and cylindrical subsystems.
We find that, for both the subsystems, the free-fermionic contribution to the entanglement entropy $S_E$ 
exhibits signatures of the phase transitions between the gapless and gapped phases.
However within the gapless phase, we find that $S_E$ does not show an expected monotonic behaviour as a function 
of the coupling $J_z$ between the suitably defined one-dimensional chains for either geometry; moreover the system 
generically reaches a point of minimum entanglement within the 
gapless phase before the entanglement saturates or increases again until the gapped phase is reached.  
This may be attributed to the onset of gapless modes in the bulk spectrum and the competition between the correlation functions along various bonds. 
In the gapped phase, on the other hand, $S_E$ always 
monotonically varies with $J_z$ independent of the sub-region size or shape.
Finally, further confirming the Li-Haldane conjecture, we find that the Schmidt gap $\Delta$ defined from the entanglement spectrum also signals the 
topological transitions but only if there are corresponding zero energy Majorana edge states that 
simultaneously appear or disappear across the transitions. We analytically corroborate some
of our results on entanglement entropy, Schmidt gap, and the bulk-edge correspondence using perturbation theory. 
\end{abstract}

\date{\today}

\maketitle
\section{Introduction}
\begin{figure}
\includegraphics[width=0.4\textwidth]{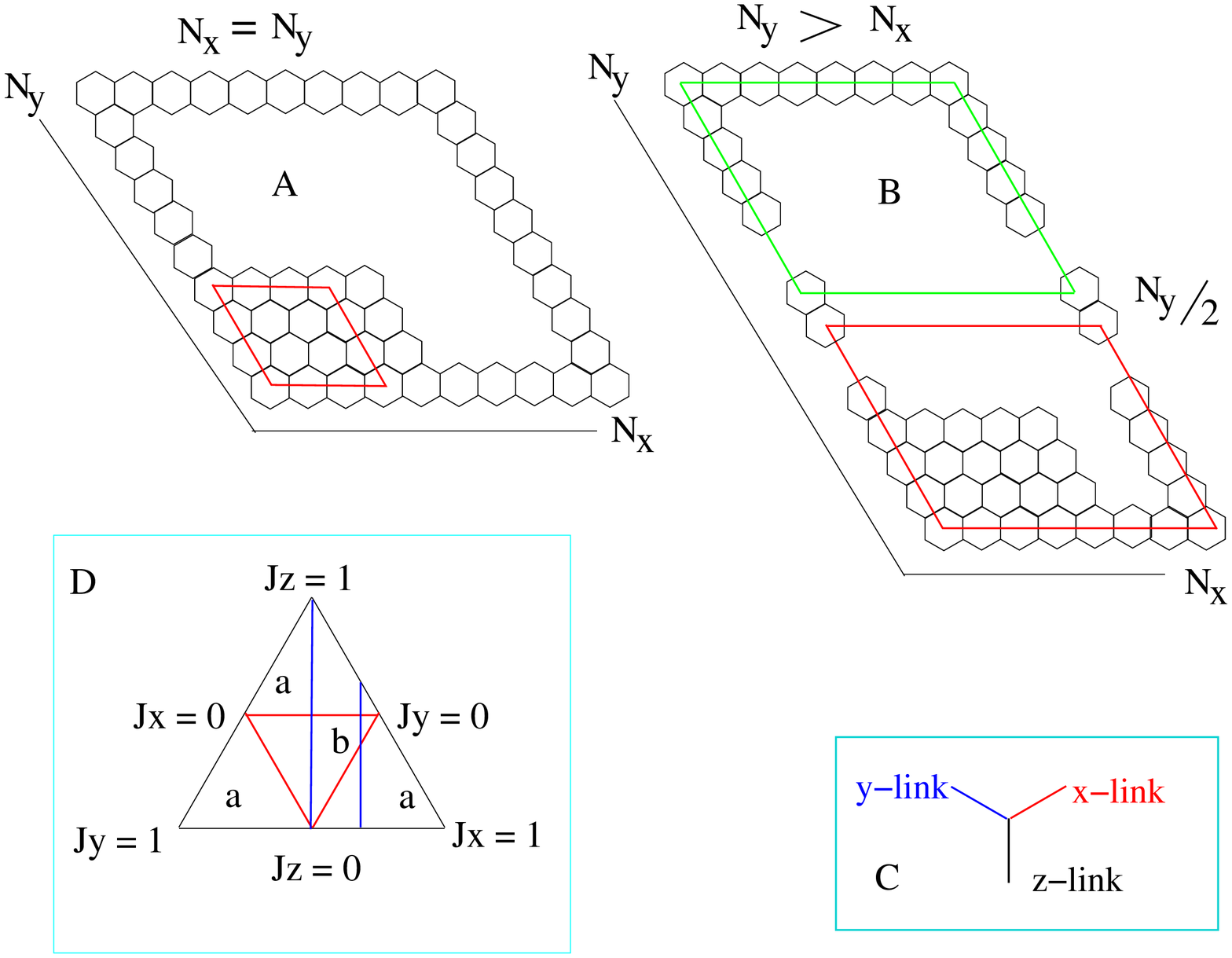}
\caption{(Color online) Top panel: geometry of subsystems chosen with square/rectangular block (plot A) and cylindrical/half-region (plot B); 
periodic boundary conditions are imposed on the full system so that a torus is formed. While considering the entanglement entropy and 
gap for the square block, we have chosen the torus such that $N_x = N_y$; on the other hand for the half-region $N_y > N_x$. Bottom panel: 
the real space links of the honeycomb lattice that correspond to the $J_x, J_y, J_z$ couplings (plot C); 
schematic phase diagram of the Kitaev model showing the three gapped phases surrounding the central gapless phase, with the vertical lines 
illustrating the contours which we study in this paper (plot D).
}
\label{Fig1}
\end{figure}

Entanglement of quantum states is a non-local phenomenon and  
is a manifestation of the superposition principle of quantum states. 
In a quantum many-body system entanglement of the constituent particles or states essentially dictates its
macroscopic properties, \textit{viz.} superconductivity, superfluidity, quantum phase
transitions etc. Recently there has been an active interest to characterize and quantify
the notion of entanglement due to its fundamental applicability in quantum information
theory, topological quantum computations, and also to numerous physical systems such as
black hole  and generic quantum many body systems ~\cite{Luigi-Amico, Eisert, 
Nishiyoka-2009}. The usual method to measure the entanglement of a part (called subsystem)
of a composite system (called full system) is to examine the reduced density matrix 
of the subsystem. To this end many definitions have been introduced for characterizing 
the entanglement \textit{viz.}  von-Neumann entropy, R\'{e}nyi entropy 
~\cite{Renyi-1960, Renyi-1965}, entanglement spectrum, Schmidt gap ~\cite{haldane-2008, ronny-2010, ronny-2010-1, konik-2013, berkovits-2015}, and suchlike. \\
These different measures of entanglement are seen to characterize physical 
systems according to their universality classes, follow certain scaling laws, and also 
detect topological orders.  For example, entanglement entropy of one-dimensional critical 
systems with size $L$ is found to vary $\sim$ $\log L$ ~\cite{calabrese-2008,calabrese-2004}.
For gapped systems in one dimension scaling of the predictions of the massive field theory 
~\cite{calabrese-2004} has also been recently verified in dimerized free fermionic models
hosting topological phases ~\cite{maiti-2014} and in spin systems ~\cite{levi-2013}.
In two and higher dimensions bosons and fermions follows different
area law. ~\cite{bombelli-1986,plenio-2005,wolf-2006}. Entanglement entropy has also been 
found to characterize quantum phase transitions ~\cite{cirac-2005,latorre-2005}.
Furthermore it has been found that for two dimensions and above entanglement entropy can characterize 
the topological properties of the system as well 
~\cite{kitaev-topo, levin-topo}. For non interacting systems, there are well defined prescriptions 
to compute the various entanglement measures discussed above; however, investigation of entanglement
properties for interacting systems are severely challenged by the complexity and the 
size of the Hilbert space as the system size is increased. Evaluating
the entanglement entropy is not straightforward in general and powerful numerical algorithms exist 
mostly for one-dimensional systems ~\cite{cirac-2004,murg-2005,zwolak-2004,g-vidal-2004,sirker-2012}. \\
This motivates us to investigate entanglement entropy in an  exactly solvable 2D quantum spin 
model, namely the Kitaev model \cite{kitaev-2003, kitaev-2006} which has attracted a lot of attention 
from researchers in condensed matter, quantum information, and specific high energy 
theorists alike.
The model realizes an unusual quantum spin-liquid ground state with short-range 
and bond-dependent correlation functions ~\cite{smandal-prl}, topological degeneracies for all 
eigenstates \cite{smandal}, and displays a tunable phase transition from a topologically trivial phase 
to a topologically nontrivial phase as the parameters of the model Hamiltonian are varied. It is known that this 
model reduces to a problem of noninteracting Majorana fermions hopping in the presence of a background 
$Z_2$ gauge field; this reduces an otherwise quartic fermionic interaction to an effective quadratic 
fermionic interactions exactly \cite{kitaev-2006}. All of these intriguing facts motivated a series 
of important studies taking the Kitaev model as a test-bed for understanding many fundamental theoretical 
concepts such as quenching and defect production \cite{krishnendu-quench-2008}, phase transitions 
\cite{vidal1-2008}, braiding statistics \cite{delgado-2007,saptarshi-2014}, dynamics of hole vacancies 
\cite{gabor-2014}, to name a select few. \\
Studies of the entanglement entropy in a particular limit of this model were undertaken early on \cite{zanardi-2005, castelnovo-2007}. 
Zanardi and coauthors verified the area law for the model and by taking 
different partitions of spin configurations had argued that the entanglement entropy can be used as 
a probe to detect topological order. However this work lacks a detailed analysis of possible subsystem 
configurations quite possibly due to the numerical complexities associated with a quantum spin model. 
This issue has been recently circumvented in an interesting work by 
Yao and Qi \cite{yao-qi2010} where they have shown that for the Kitaev model 
the entanglement entropy of a given subsystem can be separated into two parts: one part due 
to the $Z_2$ gauge fields and the other due to the free Majorana fermions. Following Yao and Qi's work,
 in this paper, we calculate the eigenvalues of the reduced density matrix for a system of free 
Majorana fermions in the two dimensional Kitaev model,  
and investigate the entanglement properties in detail for this system which have not been reported before. 
Moreover for the \textit{extended} Kitaev model, a recent density matrix renormalization group study concluded that the 
entanglement entropy and Schmidt gap may only be occasionally employed as a good indicator of the phase transitions between the various phases harboured in the 
extended system, such as the Kitaev spin-liquid phase and certain magnetic phases \cite{shinjo-2015}.
\\
In this study, we  present results of the entanglement entropy and Schmidt gap for subsystems 
with square or rectangular block geometry, and one half of the torus (herein called the half-region). For a 
square/rectangular block subsystem, the full system is a torus of size $N_x \times N_y$ (with $ N_x  = N_y $); 
for the half-region the dimension of the torus is taken such that  $ N_x  <  N_y $ and the half-region 
is defined by dividing the torus in the y-direction  
i.e choosing a length $\frac{N_y}{2}$ in $y$ direction. The subsystems under consideration are sketched in the upper panels of Fig.\ref{Fig1}.\\
 We first consider the entanglement properties for a coupled chain system
 in the Kitaev model following the geometrical setup adopted in the study of a transverse field Ising model ~\cite{konik-2013}.
 We view our 2D system as $N_y$ coupled one-dimensional periodic chain and
 consider the subsystem as the first $N_y/2$ chains. The interchain coupling between two neighbouring chain is given by $J_z$.
Also each one-dimensional chain is characterized by alternating bond interaction parameter $J_x$ and  $J_y$. \\
Let us briefly summarize the phases and physics of this model \cite{kitaev-2003, kitaev-2006}, as pertinent for understanding our results.
For small values of $J_z$, i.e for weakly coupled chain limit, the system is gapless only at the 
point $J_x=J_y$ and gapped for $J_x \neq J_y$. 
For $ J_x \neq J_y$, as $J_z$ is increased the system enters into a gapless phase for some critical value of $J_z$. However when $J_z$ exceeds another critical value, the system again enters 
into another gapped phase characterized
by large values of $J_z$. This limit is usually known as Toric code limit. The condition for gapless phase is $|J_x| \leq |J_y| + |J_z| $ and cyclic combinations of $
J_x, J_y,$ and $J_z$. For simplicity in our study we have taken all the coupling parameters to be positive, although the results presented here do not depend on the sign of the coupling parameters.  
We note that all the gapped phases are topologically equivalent: their ground state degeneracies and excitations are of the same topological nature. With this background let us summarize our results.\\
 \textit{Half-region geometry}: For the Kitaev model in the half-region geometry the entanglement entropy $S_E$ grossly increases as $J_z$ is increased, until in the Toric code limit 
 it decreases and saturates to a finite value in the large $J_z$ limit. 
 However within the gapless phase there is a nonmonotonic dependence of $S_E$ on the strength of $J_z$, manifesting itself as oscillations;
 the height and number of these oscillations within the gapless phase are dependent on the system sizes chosen. At the gapless$-$gapped transitions a cusp in $S_E$ is generically visible.
 The Schmidt gap $\Delta$ in this subregion shows a first order jump from a finite value in the gapped weakly-coupled chain limit  
 to zero in the gapless phase; finite size effects for $\Delta$ are almost absent for this topological phase transition.
 This is in contrast to the case of the 2D transverse field Ising model on a similar geometry 
 where $\Delta$ displays logarithmic scaling with system size $\sim \textrm{const.}/ \ln{(N_y/\pi)}$ ~\cite{konik-2013}.
 Moreover as we further increase the inter-coupling chain and reach the large $J_z$ gapped regime i.e the 
 Toric code limit, the entanglement gap \textit{still} remains zero. Thus we find that although the 
 two gapped phases are topologically identical, the Schmidt gaps of a given subsystem are different in 
 the two gapped phases indicating that other properties of the subsystem compared with the full system 
 play an important role. In particular we ascribe this to the presence or absence of Majorana edge states in the two gapped phases of the system 
 and infer that a vanishing Schmidt gap does not necessarily imply topological order.\\
\textit{Square/Rectangular block geometry}: In this case the qualitative behaviour of the entanglement entropy 
 depends crucially on the details of the connectivity of the system; nevertheless a nonmonotonic behaviour of $S_E$ persists within the gapless phase here as well.
The cusps in $S_E$ are conspicuous, as in the half-region, at the phase transitions.
 The Schmidt gap is vanishingly small within the gapless phase and shows minute nonmonotonic variations, especially close to the conformal critical point; 
 in the gapped phases, where the entanglement entropy satisfies the area law  i.e  $(S \sim \tilde{\alpha} L^{d-1})$, these variations are proportionally washed away 
 and the coefficient $\tilde{\alpha}$ depends nontrivially on the underlying system parameters.\\
The organization of the rest of the paper is as follows. In Sec.\ref{section-1}, we further explicate 
on the physics and phase diagram of the Kitaev model.
We also outline the formalism (following Ref. \onlinecite{peschel-2003}) that we employ to investigate the
entanglement properties of this model. In Sec.\ref{section-2}, 
we present the results of the entanglement entropy and Schmidt gap for the half-region which 
has a cylindrical geometry; in Sec.\ref{section-3} we discuss the case of square and rectangular blocks. 
In Sec.\ref{section-4} we present some analytical 
perturbative calculations to corroborate some of our numerical findings presented in Secs.\ref{section-2} 
and \ref{section-3}.  We summarize our primary results in Sec.\ref{section-5}.

\section{Kitaev model and entanglement properties}
\label{section-1}
Due to the recent interest in the Kitaev model, a vast literature is already available on
various aspects of this system. However, in this section, we introduce the model briefly for 
completeness and self-sufficiency of the article. The original model is defined on a hexagonal 
lattice with each site associated with a spin $\frac{1}{2}$  object. 
Each spin interacts with its nearest-neighbour and the coupling strength depends on the 
directionality which is in contrast to the Heisenberg interaction ~$\vec{s}_i \ldotp \vec{s}_j$.
In the hexagonal lattice, 
there are three different orientations of the bonds and we label them as "$x$-bonds", "$y$-bonds" 
and "$z$-bonds". Two nearest-neighbour spins joined by an $\alpha$ ($\alpha =x,y,z $) bond 
interact with the $\alpha$ component of their spins contributing a term $\sigma^\alpha_j\,\sigma^\alpha_k$ 
to the Hamiltonian. Here $j$ and $k$ represent the site indices for two nearest-neighbour spins 
situated at the two ends of the bond. The model Hamiltonian ~\cite{kitaev-2006} is 
given by

\begin{multline}
H = - \sum_{<j,k>_{\alpha}}J_{\alpha_{jk}}\sigma^{\alpha}_j\,\sigma^{\alpha}_k,~~~~(\alpha=x,~y,~z),
\label{spin-hamiltonian}
\end{multline}

 where, $J_\alpha$'s are dimensionless coupling constants, $\sigma^{\alpha}_k$ is the $\alpha$-component of the Pauli 
matrices. Following  Ref. ~\onlinecite{kitaev-2006}, we introduce a set of four Majorana fermions $\{ b^x_k, b^y_k, b^z_k, c_k \}$ at a given site 
`$k$' to represent the Pauli operators. Notice that this definition implies that the spin operators live in an enlarged Hilbert space. 
We define the Pauli operators in this enlarged Hilbert space as, $\tilde{\sigma}_k^{\alpha}= i b^{\alpha}_k c^{\phantom{\dagger}}_k$ ($\alpha=x,y,z$). 
We have used $\tilde{\sigma}$ instead of $\sigma$ to denote the fact that these operators are defined in an enlarged Hilbert space. To get the physical spin operators $\sigma$
one has  to enforce the projection in the physical Hilbert space using the operator  
$D_k=  b^{x}_k b^y_k b^z_k c_k=1$, at a given site `$k$'.  Substituting the above definition, 
we can rewrite the  Hamiltonian in Eq.(\ref{spin-hamiltonian}) (in the enlarged Hilbert space) as
\bea
\tilde{H} &=&  \frac{i}{2} \sum_{< j,k >_{\alpha}} J_{\alpha_{jk}} \hat{u}_{jk} c_j c_k,
\label{majorana-hamiltonian}
\eea

where $ \hat{u}_{jk}=i b^{\alpha_{jk}}_j b^{\alpha_{jk}}_k $ are the link operators  defined on a given link $<jk>$.
The remarkable fact which makes the Kitaev model an integrable system is that these $ \hat{u}_{jk}$ operators defined
on each link mutually commute with each other and also commute with the Hamiltonian in Eq.(\ref{majorana-hamiltonian}). 
They play the role of  static $Z_2$ gauge field operators as one may readily check that $ \hat{u}^2_{jk}=1$. 
Here we have  $\alpha_{jk} = x, y, z$ depending on whether the `$j$' and `$k$' indices form an $x, y$ or $z$ link.  \\
The original Hamiltonian as given in Eq.(\ref{spin-hamiltonian}) has been transformed into an equivalent problem as described by 
Eq.(\ref{majorana-hamiltonian}); now one has to solve for a free Majorana fermion hopping problem in the presence of
static $Z_2$ gauge fields. For the detailed solution and the phase diagram we refer the reader to the original work 
by A. Kitaev ~\cite{kitaev-2006}; a pictorial summary is presented in the lower panel of Fig.\ref{Fig1}.  
Working in the extended Hilbert space of the spin operators, a general eigenstate 
of the  Hamiltonian in Eq.(\ref{majorana-hamiltonian}) can be written as $ | \tilde{\Psi} \rangle =  |\phi(u) \rangle |u \rangle$ \cite{kitaev-2006}. 
$|\phi(u) \rangle$ is obtained 
by considering the Hamiltonian identical to Eq.(\ref{majorana-hamiltonian}) with $\hat{u}_{jk}$ replaced
 by its eigenvalues $u_{ij} = \pm 1$. $|u \rangle$ describes the gauge field configurations. The actual eigenstate  $| \Psi \rangle$ belonging to the physical Hilbert space is then obtained by 
 projecting $| \tilde{\Psi} \rangle$ onto the physical Hilbert space and is given by \cite{yao-qi2010}
\begin{eqnarray}
| \Psi \rangle = \frac{1}{\sqrt{2^{N+1}}} \sum_g \mathcal{D}_g |\phi(u) \rangle |u \rangle,
\end{eqnarray}

where $N$ is the total number of sites present in the system. $\mathcal{D}_g= \prod_k D_k $ where $g$ denotes a set of sites  and the product runs over   all the sites `$k$' within a  given set `$g$'. 
The sum over `$g$' is over $2^{N}$ possible combination of sets.  The sum $\frac{1}{\sqrt{2^{N+1}}} \sum_g \mathcal{D}_g$  represents the gauge average 
over equivalent copies in the extended Hilbert space.  In Yao and Qi's work ~\cite{yao-qi2010} it has been shown that due to the 
special structure of the Hamiltonian in Eq.(\ref{majorana-hamiltonian}), the  entanglement entropy ($S_A$) of a 
subsystem `$A$' for a given eigenstate $| \Psi \rangle$ consists of two contributions and may be written as
\bea
\label{gauge-fermi}
S_A=  S_{A,F} + S_{A,G} -\rm{ log } 2.
\eea

In the above expression, the contribution $S_{A,G} $ comes exclusively from the $Z_2$ gauge fields and equals  
$L~{\rm log }2$  when the subsystem shares  $L$ number of bonds with the rest of the system. The quantity 
$-\rm{log\;} 2$ is a topological quantity 
known as the topological entanglement entropy which is same for both the gapless 
and gapped phases of the Kitaev model.  
$S_{A,F}$  is the contribution from the free Majorana fermions and is obtained from the reduced density matrix $ \rho_{A,F} ={\rm{Tr_{B}}} |\phi(u) \rangle \langle \phi(u) |$ as 
$S_{A,F} = -\textrm{Tr}[ \rho_{A,F}\log{\rho_{A,F}} ]$. 
In this article we calculate $S_{A,F}$ for the ground state sector where the product of $u_{ij}$ over a plaquette is 1 i.e. the vortex-free sector; 
indeed $S_{A,F}$ is a gauge independent quantity.  
Thus the Hamiltonian which is central to our investigation is obtained from Eq.(\ref{majorana-hamiltonian}) by replacing $u_{ij}=1$ and is given by, 
\bea
H^{\prime} &=&  \frac{i}{2} \sum_{\langle j,k \rangle} J_{\alpha_{jk}}  c_j c_k.
\label{majorana-hamiltonian-1}
\eea
 
For a subsystem with boundary of linear size $L$,  
it can been shown \cite{yao-qi2010, zanardi-2005} that in the limit $L \rightarrow \infty$, an area law~\cite{Luigi-Amico} $ S_{A,F} \sim \tilde{\alpha} L$ holds, 
where $\tilde{\alpha}$ is a 
non-universal constant that contains necessary information of the entanglement between the subsystem and the rest. \\
We note here, that the entanglement entropy can be written as Eq.(\ref{gauge-fermi}) for any shape and size of the subsystem and 
thus the entanglement entropy of a given subsystem can in principle be calculated by applying the formalism of Ref. \onlinecite{peschel-2003}. 
As mentioned earlier, one of the main objective of this article is to investigate
the behaviour of $\tilde{\alpha}$ in the  parameter space of the Kitaev model. 
Specifically, we would like to know, apart from an area law, what additional information might be extracted from the entanglement entropy of the system.
For this we examine the entanglement entropy and entanglement gap as a function of the model parameters $J_x,J_y,J_z$. 
For simplicity we choose certain particular contours in the phase space of Kitaev model to ascertain how the entanglement property 
varies along these paths. 
The free Majorana fermion hopping Hamiltonian given by Eq.(\ref{majorana-hamiltonian-1}) has two different topological phases, 
gapless and gapped phases as shown in lower panel of Fig.\ref{Fig1}. 
This Hamiltonian can easily be diagonalized by a Fourier transformation and the 
ground state correlation functions can be subsequently calculated. Following Ref.~\onlinecite{peschel-2003} this now reduces 
the calculation of the von-Neumann entanglement entropy of a given subsystem `$A$' to simply calculating the two-point correlation functions 
 in order to obtain the eigenvalues of the 
reduced density matrix; this is described as follows \cite{kitaev-2006, peschel-2003}.\\
The correlation matrix of the full system has elements given by
\begin{equation}
P_{ij} = \langle \Psi| c_jc_i |\Psi \rangle - \delta_{ij}.
\end{equation}
This can be rewritten as 
\bea
P = iQ
\begin{bmatrix}
0  & -1     &       &        &      &   \\
1  & 0 & \ddots & \ddots &        &  \\
  &        & \ddots & \ddots & 0 & -1 \\
 &       &        &      & 1      & 0
\end{bmatrix}
Q^{T},
\eea

where $Q$ is the matrix whose columns are composed alternatively of the real and imaginary parts of the eigenvectors of the Majorana Hamiltonian. 
The eigenvalues $\lambda_1,...,\lambda_i,...\lambda_{N_A}$ of $\tilde{P} \subset P$ lie between $[-1,1]$, where $\tilde{P}$ denotes the correlation matrix restricted to the subsystem and 
$N_A$ denotes the total number of sites inside the subsystem; then the eigenvalues of the reduced density matrix obtained from $| \phi(u) \rangle$ can be written as \cite{calabrese-2010,vidal-2004},

\bea
\Gamma(s_1,...s_i,...s_{N_A})= \prod_{i=1,N_A} \frac{1+ s_i \lambda_i}{2}.
\label{density-eigvl}
\eea

$s_i$ can take values $\pm 1$ and thus we obtain in total $2^{N_A}$ eigenvalues of the density matrix. The von-Neumann entanglement entropy is thus given by ~\cite{yao-qi2010},
\bea
\label{final-ent}
S_{A,F} = \sum^{N_A}_{i=1} \frac{1+ \lambda_i}{2} {\rm log} \frac{1+ \lambda_i}{2} + \frac{1- \lambda_i}{2} {\rm log} \frac{1- \lambda_i}{2}.
\eea  
 
Entanglement gap or Schmidt gap is another useful quantity to characterize the entanglement properties; in particular this gap has been conjectured to be capable of detecting topological order 
\cite{haldane-2008} and phase transitions \cite{konik-2013}.
 The entanglement gap may be defined as ~\cite{haldane-2008, konik-2013}
\bea
\Delta_A= - (\rm{log} \Gamma_M - \rm{log} \Gamma_M^{\prime}),
\eea
where $\Gamma_M$ and $\Gamma_M^{\prime} $ are the largest and the second largest eigenvalues of the reduced density matrix.

From Eq.(\ref{density-eigvl}), it is clear that $\Gamma_M$ is obtained when the contribution $\frac{1+ s_i \lambda_i}{2}$ 
is maximized i.e one has to consider the contribution $\textrm{Max} \Big(\frac{1+ \lambda_i}{2}, \frac{1- \lambda_i}{2}\Big)$ for a 
given $\lambda_i$. The second largest eigenvalue is obtained by changing  $(1+ \lambda)/2$ to $(1-\lambda)/2$ for the  smallest 
value of $|\lambda|$. With this, entanglement gap can be simplified to
\bea
\label{final-gap}
\Delta_A= {\rm log} \frac{(1+ |\lambda|_{min})}{(1- |\lambda|_{min})}.
\eea
We evaluate the quantities given by Eq.(\ref{final-ent}) and Eq.(\ref{final-gap}) numerically and discuss the results 
in Sec.\ref{section-2}. \\
Before we end this section we show how the spectrum, eigenvalues and eigenvectors of the Hamiltonian given in Eq.(\ref{majorana-hamiltonian-1}) may be constructed, required for our analytical computations in Sec. \ref{section-5}. 
This Hamiltonian can be easily 
diagonalised by using a Fourier transformation, $c_j= \frac{1}{\sqrt{N}} \sum_{\vec{k}} e^{i \vec{k} \cdot \vec{r}_j} c^{\phantom{\dagger}}_{\vec{k}}$, with $c_{\vec{k}}=c^{\dagger}_{-\vec{k}}$ 
which ensures that $c_j$'s are Majorana fermions. The Hamiltonian then reduces to
\begin{eqnarray}
&& H^{'}= \sum_{k_x,k_y \geq 0} i f(\vec{k}) c^{\dagger}_{\vec{k},a} c^{\phantom{\dagger}}_{\vec{k},b} + {\rm h.c},\\
&& H^{'}=\sum_{k_x,k_y \geq 0} \psi^{\dagger}_{\vec{k}} \tilde{h}_k \psi^{\phantom{\dagger}}_{\vec{k}}, \psi^{\dagger}_{\vec{k}}= (c^{\dagger}_{\vec{k},a}, c^{\phantom{\dagger}}_{\vec{k},b}), \\
&& \tilde{h}_k= \left( \begin{array}{cc}
 0&if(\vec{k}) \\
-if^*(\vec{k})&0 \end{array} \right).
\label{kspace-hamiltonian}
\end{eqnarray}
$a$ and $b$ correspond to the two sublattice points. The dispersion $f(\vec{k})$ is given by
\begin{equation}
\label{eq:dispersion}
|f(\vec{k})|= J_x e^{i  k_x} + J_y e^{i k_y} + J_z,
\end{equation}
with $k_x$ and $k_y$ being the components of $\vec{k}$
along the $x$ and $y$ bonds respectively. $k_x$ and $k_y$ can be rewritten as the Cartesian components of $\vec{k}$
along $x$ and $y$ axes as $k_x=\frac{q_x}{2} + \frac{\sqrt{3} q_y}{2}$ and  $k_y= -\frac{q_x}{2} + \frac{\sqrt{3} q_y}{2}$.
With the unitary transformations 
\begin{eqnarray}
c_{\vec{k},a}&=&\frac{i \exp{i \theta_{\vec{k}}}}{\sqrt{2}}(\alpha_{\vec{k}}+\beta_{\vec{k}}), \nonumber \\
c_{\vec{k},b}&=&\frac{1}{\sqrt{2}}(\alpha_{\vec{k}} - \beta_{\vec{k}}),
\end{eqnarray}
the Hamiltonian in Eq.(\ref{kspace-hamiltonian}) may be diagonalised as 
\begin{eqnarray}
H^{'}= \sum_{k_x,k_y \geq 0} |f(\vec{k})| \Big( \alpha^{\dagger}_{\vec{k}} \alpha_{\vec{k}} -  \beta^{\dagger}_{\vec{k}} \beta_{\vec{k}} \Big).
\end{eqnarray}
$c_{\vec{k}}$'s are defined in the first half of the Brillouin zone (HBZ) and 
$e^{i \theta_{\vec{k}}}= \frac{f(\vec{k})}{|f(\vec{k})|}$. The ground state $|\mathcal{G} \rangle$ is obtained by filling 
all the $\beta_{\vec{k}}$ modes,  i.e, $|\mathcal{G} \rangle= \prod_{\vec{k} \in HBZ} \beta^{\dagger}_{\vec{k}} |0 \rangle$
where $|0 \rangle$ denotes the fermionic vacuum. \\
%
The discretization of $k_x$ and $k_y$ employed here for a finite size lattice is as follows. For the hexagonal lattice unit vectors are chosen as
$\vec{a}_1= \vec{e}_x,~~ \vec{a}_2= \frac{1}{2} \vec{e}_x + \frac{\sqrt{3}}{2}  \vec{e}_y $. These yield the
reciprocal vectors as $\vec{G}_1=2 \pi \left( \vec{e}_x -\frac{1}{\sqrt{3}} \vec{e}_y \right),~~~\vec{G}_2= \frac{4 \pi}{\sqrt{3}} \vec{e}_y$. Any $\vec{k}$ vector 
is discretized as $\vec{k}= \frac{n_x}{N_x} \vec{G}_1 + \frac{n_y}{N_y} \vec{G}_2$. 
$k_{x/y}= \vec{k} \cdot \vec{a}_{x/y}$ with $\vec{a}_{x/y}= \pm \frac{1}{2} \vec{e}_x + \frac{\sqrt{3}}{2} \vec{e}_y$.
Here $\vec{a}_{x/y}$ are the unit vectors joining two z-bonds by $x$ and $y$ interactions respectively, and $\vec{e}_{x/y}$ are the unit vectors along $x$ and $y$-directions respectively. 
Therefore one obtains $k_x= \frac{2 \pi n_y}{N_y}$ and $k_y=-\frac{2 \pi n_x}{N_x} + \frac{2 \pi n_y}{N_y}$; 
$N_x$ and $N_y$ denote the number of unit cells taken in the $x$ and $y$-direction 
respectively, and $n_{x/y}$ varies from $1$ to $N_{x/y}$.
\begin{figure*}
\includegraphics[width=18cm]{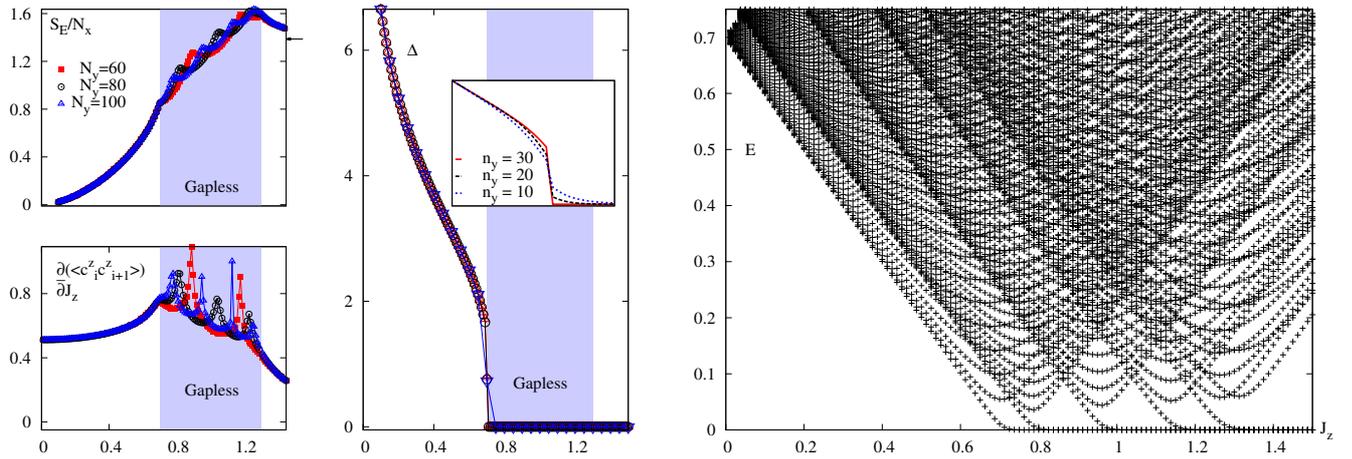}
\caption{(Colour online) Entanglement properties of cylindrical subregions for $J_y/J_x=0.3$ as a function of $J_z$ in the Kitaev system; gapless phase is shaded blue. 
Left panel: top plot shows the entanglement entropy for various 
sizes of the full system with the aspect ratio fixed at $N_y/N_x = 10$, with the arrow denoting the saturation value in the large-$J_z$ limit; 
bottom plot shows the derivative of correlation functions obtained analytically, with the oscillations present at 
the same values of $J_z$ as the entanglement entropy, the former resulting in the latter. 
Middle panel: Schmidt gap for the half-region is plotted against $J_z$. Initially when the 
 system is in the gapped phase, the entanglement gap gradually decreases as $J_z$ increases, 
 but it goes to zero in a first order jump when it enters the gapless regime. However the gap remains zero when 
 $J_z$ is increased further and the system transits to the large $J_z$ gapped phase, consistent with zero energy edge-modes in the latter two phases (right panel); 
 inset shows the variation of the gap for different number of chains in the subregion $n_y$ as indicated. We have taken
 a system with $N_x=6, N_y=60$. The gap is a generic feature of a subsystem comprising any number of parallel chains. 
 However the first order jump becomes sharper as $n_y$ is increased.
 Right panel: positive spectrum of the Majorana Hamiltonian for the half-region with $J_y/J_x=0.3$ as a function of $J_z$ for a $10 \times 100$ system. 
 Energy modes become zero energy modes at the transition from the gapped to gapless phase; we have checked that these are indeed edge states localized on the boundary. However in 
 the large $J_z$ limit where the system again becomes gapped, the edge modes persist and remain gapless.
}
\label{Fig2}
\end{figure*}
\section{Entanglement in cylindrical subregions}
\label{section-2}
In this section we discuss the entanglement entropy and gap for the half-region which has a cylindrical geometry.
Here we have taken a torus of $N_x \times N_y$ unit cell which can be thought 
of as $N_y$ coupled one-dimensional chains where each chain has $N_x$ sites such that 
$N_x\neq N_y$ \cite{konik-2013}. We have kept the aspect 
ratio $N_y/N_x=10$; our qualitative results are independent of this ratio. 
The subsystem is defined such that it contains $N_y/2$ coupled one-dimensional 
chains. This is because the system is periodic in the $x$-direction, hence one can employ a Fourier transform in the $x$-direction 
which reduces the entanglement study to that of decoupled one-dimensional chains for each $k_x$-mode in $y$-direction. 
We consider the case $J_y/J_x=0.3$, such that $J_x=1.0$ and $J_y=0.3$. Thus, in the plots of entanglement properties 
as shown in Fig. 2 as a function of $J_z$, $J_z$ is also scaled in units of $J_x$.
In the upper left panel of Fig.\ref{Fig2}, we have plotted the entanglement entropy for the half-region as a function of $J_z$. 
It may be worthwhile to note 
that a similar region on a torus had previously been considered for studying the entanglement properties 
~\cite{yao-qi2010}, where the entanglement entropy was studied for each $k_x$-mode for a given set 
of parameter values in the gapless and gapped regime; however that work lacked a comprehensive analysis of the 
entanglement entropy, in particular the physics contained in the non-universal parameter $\tilde{\alpha}$ appearing in the area law 
along with its variation with the system parameters, which we provide in this work. \\
As may be seen from the plot, for each of these systems a cusp is prominent at $J_z=J_x+J_y$,  
which is the transition point from the gapless to the gapped phase; moreover within the gapped phases the area law is clearly validated. 
However, inside the gapless phase, the entanglement entropy has intermittent peaks whose number increases with increasing system size for 
a given aspect ratio. The entanglement entropy for the half-region with $J_x=J_y$ (not shown) is similar to that 
obtained for $J_x \neq J_y$ with the initial monotonic increase for $J_x \neq J_y$ being absent. We note  
that the intermittent peaks we find in the gapless phase for the entanglement entropy is very similar 
to results for the fidelity-susceptibility~\cite{gu-2008}. The number of peaks  
is found to increase linearly with the number of chains in the total subsystem. \\ 
We now provide a possible explanation for the oscillations of the entanglement entropy in this half-region. 
We know that the fundamental quantity which determines the entanglement entropy in a free-fermion system 
is the two point correlation function \cite{peschel-2003}. 
To elucidate the relation of the oscillations present in the entanglement entropy with the two-point correlation functions, 
we consider the $z-z$ nearest-neighbour correlation function $\mathcal{C}_{zz} \equiv \langle c^z_ic^z_{i+1}\rangle$.
The expectation value is taken in the half-filled Majorana system
for many different system sizes keeping the aspect ratio identical in accordance with the upper left panel of Fig.\ref{Fig2}.
This may be seen to be expressed analytically as
\begin{eqnarray}
&&\mathcal{C}_{zz}=\sum_k \cos \theta_k=\frac{Re(f(\vec{k}))}{|f(\vec{k})|},
\label{eq:corr1}
\end{eqnarray}
where 
$|f(\vec{k})|$ is the dispersion of the Kitaev model given in Eq.(\ref{eq:dispersion}).
In the lower left panel of  Fig.\ref{Fig2} we have plotted the derivative of $C_{zz}$ with respect to $J_z$, where we  observe the 
appearance of similar oscillations.
The oscillations in the correlation function produces the oscillations in the derivative of it as well, but of higher magnitude, and explain the oscillations in the entanglement 
entropy (the oscillations in the derivative of correlation function are more pronounced than the oscillations in the correlation function itself due to the presence of an additional 
factor $|f(\vec{k})|$ in the denominator which becomes vanishingly small for some values of $k_x, k_y$.). \\
The oscillations in $C_{zz}$ may also be understood alternatively from the form of Eq.(\ref{eq:corr1}). In the gapless phase the dispersion $f(\vec{k})$ vanishes at 
certain $\vec{k}$ points in the Brillouin zone; 
however due to the discrete numerical evaluation of $\vec{k}$-points, the dispersion may take values that are arbitrarily close to zero depending on the system size 
and the value of the model parameters. This can occur only in the absence of a bulk gap, eventually 
resulting in the oscillations of the entanglement entropy, the magnitude of which depends  
on the full system size; we point out that this is not a violation of the area law, the fulfilment of which 
is more conspicuously visible in the gapped phase for the same reason.  \\

Let us turn now to the entanglement gap. In the middle panel of Fig.\ref{Fig2}, we have plotted the Schmidt gap for the half-region for $J_x \neq J_y$. We have found that 
as long as we are in the weakly coupled chain limit and in the gapped phase ($J_{x(y)} \ge J_{y(x)} + J_z $) the Schmidt gap is finite, but it 
abruptly goes to zero when it enters the gapless region (when  $J_i \leq J_j + J_k$ where $i,~j,~k$ could be $x~,y~,z$ and its cyclic combinations). 
The gap continues to remain zero even in the large 
$J_z$ limit as the system enters the other gapped regime (when $J_z \geq J_x + J_y$). This is a new result and
we conclude that the zero Schimdt gap is not a sufficient condition to signal the presence of a gapless phase as was done in an earlier study \cite{yao-qi2010}. 
To ascertain the origin of the zero gap, we have investigated the edge mode spectrum of the half-region because 
the presence of topological order and edge states are intimately linked to the entanglement spectrum \cite{haldane-2008,lukasz-2010}. 
 We have found that for the half-region geometry, which has zig-zag $x-y$ chains at the ends, the system 
harbours gapless edge modes in the large $J_z$ phase as well 
as in the gapless phase. This has been shown in the right panel of Fig.\ref{Fig2}. However 
the system has a gapped edge spectrum in the weakly coupled gapped phase. Our findings of the Majorana edge mode spectrum  
agree with earlier analytical results on the same \cite{diptiman} which, together with our findings on the Schmidt gap, 
corroborates the previous observation \cite{lukasz-2010,turner} that a gapless edge state is associated with a gapless entanglement spectrum.\\
Our results for entanglement entropy and Schmidt gap may be compared with that of a previous study \cite{konik-2013}, 
where the authors studied the entanglement entropy and Schmidt gap of the two-dimensional transverse field Ising model. Both the entanglement entropy and 
the gap have different behaviour than what we have found here. The characteristic scaling of entanglement gap and the crossing 
of entanglement gap at the transition point when the system goes from gapped to gapless regime is absent in the Kitaev model. 
This may be attributed to the fact that the transition in the transverse model is continuous and for the Kitaev model
the transition from gapless to gapped one is a topological one.
Thus we can see that the entanglement gap does not depend on the gapless or gapped nature of the bulk spectrum but rather it is a property of the gapless edge mode: 
\textit{when such gapless edge modes
exists, the entanglement gap is zero}. 
However the entanglement gap can can also be zero without the presence of a gapless edge mode, for instance by breaking time reversal symmetry but preserving the inversion 
symmetry \cite{turner}, the aspect which we do not pursue here.\\
We note that  
the above behaviour of entanglement entropy and gap as presented in Fig.\ref{Fig2} is also true for any number of chains taken periodic 
in the $x$-direction. We demonstrate this in the inset of the middle panel of  
Fig.\ref{Fig2} where we have plotted the entanglement gap for different number of chains $n_y$; it is evident that 
the entanglement gap remains qualitatively unchanged except that the change from a finite gap to a zero gap becomes sharper 
as the number of chains in the subsystem increases. \\
\section{Entanglement in square/rectangular subregions}
\label{section-3}
\begin{figure*}[ttp]
\label{Rec-JxJy}
\includegraphics[width=18cm]{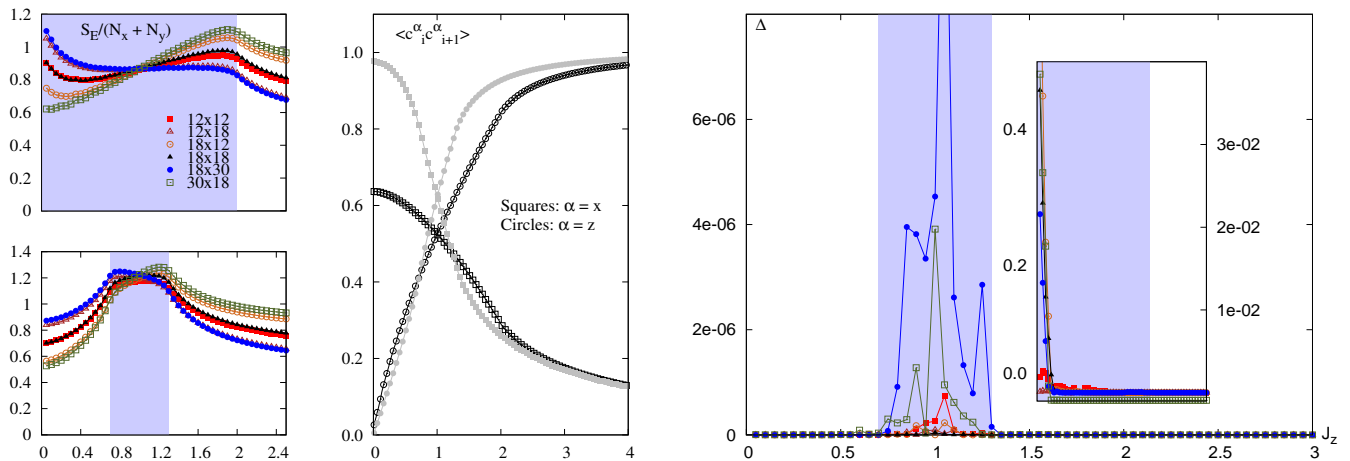}
\caption{(Colour online) Entanglement properties of general square and rectangular subregions as a function of coupling $J_z$ in a $36 \times 36$ Kitaev system; 
gapless phase is shaded blue as in Fig.\ref{Fig2}.
Left panel: Entanglement entropy of various subregions across the phase transitions for $J_y/J_x = 1$ (top) and $J_y/J_x = 0.3$ (bottom); 
at the transition to/from the gapless phase there is a cusp in the free fermion entanglement entropy. However within the gapless phase the entropy does not monotonically change with varying $J_z$, and depends 
on the transverse coupling and geometry of the subregion.
Middle panel: nearest-neighbour correlations for $x-x$ and $z-z$ bonds in the ground state for $J_y/J_x = 0.3\textrm{ (closed symbols)}, 1.0\textrm{ (open symbols)}$, 
 the nonmonotonicity of which in turn suggests the nonmonotonicity of the entanglement entropy in the left panel.
 Right panel: Schmidt gap $\Delta$ for various subsystem geometries with $J_y/J_x=0.3$. Within the gapless phase there are stronger variations in $\Delta$ than when within or 
 closer to the transition to the gapped phases. Inset displays similar results for $J_y/J_x=1$ (left $y$-axis corresponds to the largest subsystem $30 \times 18$). 
 The Majorana edge states (not shown) are not strongly localized along the 
 boundaries in this case  having finite extension in the bulk as well. Thus the putative correspondence between the gapless edge mode and the 
gapless entanglement spectrum might not be straightforward here.} 
\label{Fig3}
\end{figure*}
In this section we present numerical results for the entanglement entropy and gap of a finite block of square or rectangular geometry as 
shown in Fig.\ref{Fig1}. Firstly, we analyze the variation of the entanglement entropy and gap as we 
vary $J_z$ for fixed values of $J_x, J_y$ (i) along the contour $J_x=J_y$ when the system changes from the gapless to the gapped phase 
and (ii) along the $J_x\neq J_y$ line when the system changes from the gapped to gapless and again to gapped phase. 
And secondly, we study the variation of the entanglement entropy and gap as a function of system size and shapes, mainly from
square to rectangular geometry. 
Our primary motivation in undertaking these two approaches is to discern signatures of the transitions, individual phases, and Majorana edge states from the 
non-universal constant $\tilde{\alpha}$. \\
In the left panels of Fig.\ref{Fig3} we plot the variation of the entanglement entropy as a function of $J_z$ for various subsystem
sizes for $J_x=J_y$ (upper plot) and for $J_y/J_x=0.3$ (lower plot). For $J_y/J_x=0.3$, similar to the cylindrical subregions, 
$J_x=1.0$ and $J_y=0.3$. Thus, in the successive plots of entanglement properties for $J_y/J_x=0.3$ as shown in Fig. \ref{Fig3} 
as a function of $J_z$, $J_z$ is also scaled in units of $J_x$. The results are scaled by the length of the boundary in each case. 
In the first case we observe that initially as we increase $J_z$, the entanglement entropy decreases for
all  subsystem sizes. However  depending on the subsystem shapes, the entanglement entropy either starts to increase
after  some  value of $J_z$ (the  particular value of $J_z$ depends on the shape of the subsystem) or remains almost constant
before entering the gapped phase where the entanglement entropy decreases uniformly irrespective of subsystem sizes and shapes.
Thus we find that the entanglement entropy may be  nonmonotonic or monotonic depending on the system parameters. We believe that 
the nonmonotonic variation of the entanglement entropy 
across phase transitions and also within the given gapless phase (implying that there is point of minimal entanglement in the gapless phase 
which depends on the subregion geometry) is a fact worth highlighting. Indeed an oscillatory nonmonotonicity was also observed for the entanglement entropy in the half-region geometry 
within the gapless phase; that was however attributed to the onset of gapless $\vec{k}$ modes in the bulk. \\ 
We explain the above variation
of entanglement entropy qualitatively again with the correlation function. We note that an exact quantitative determination
of entanglement entropy requires an analytical diagonalization of correlation matrix which is beyond the scope
of any known technique to our knowledge. However we can expect that the boundary bonds shared between the subsystem and the system would 
determine qualitatively the entanglement entropy. In the middle panel of Fig.\ref{Fig3}, we have plotted the two point
Majorana fermionic correlation function for $x$-bonds and $z$-bonds. Any rectangular or square block shares $x$-bonds with the
system at left and right boundary. On the other hand it shares $z$-bonds at the upper and lower
boundary bonds. As we see from the middle panel of Fig.\ref{Fig3}, the two bond correlation functions behave differently
as we increase the value of $J_z$. As the ratio of number of $z$-bonds and $x$-bonds are different for different subsystems,
the entanglement entropy in turn shows a nonmonotonic behaviour. Thus we see that by  manipulating the ratio of length to breadth
of a rectangular subsystem, one can go from monotonic to nonmonotonic dependence of the entanglement entropy within the gapless phase.\\
For $J_y/J_x=0.3$, entanglement entropy is plotted in the lower left panel of Fig.\ref{Fig3},
 where we also observe similar behaviour; irrespective of the subsystem size, the entanglement entropy shows monotonic behaviour in the gapped phases. 
For this case, in the gapless phase, as explained in the previous paragraph and for the half-region of the previous section, 
the entanglement entropy or may not show monotonic behaviour depending on the 
values of $J_y/J_x$ and $J_z/J_x$. Indeed for a given $b_x/b_z \propto n_y/n_x$ ratio 
(where $b_x$ and $b_z$ are the number of $x$-bonds and $z$-bonds shared between a square/rectangular subsystem with the system, and $n_x \times n_y$ is the subsystem size) the 
scaled entanglement entropy curves seem to collapse on top of each other both for $J_y = J_x$ and $J_y \neq J_x$. 
This  may be explained qualitatively as follows: the primary contribution to the entanglement entropy comes from the nearest-neighbour bond-correlation between the system and the subsystem.
 Then let $S(n_x, n_y)$  denote the entanglement entropy of a certain rectangular subsystem geometry having a total of $b_x (\propto n_y)$ and $b_z (\propto n_x)$  boundary $x$-bonds and $z$-bonds respectively. 
 For simplicity we may assume $S(n_x,n_y)= \kappa_x n_y + \kappa_z n_x$ where $\kappa_x$ and $\kappa_z$ are the contributions due to a boundary  $x$-bond  and $z$-bond; 
 then $S(n_x,n_y)/(n_x+ n_y)$ takes on a given value for a given $n_x/n_y$. 
 However deviations from this qualitative argument are expected as contributions come from all possible correlation functions  beyond the 
 nearest-neighbour correlations. \\
Now we turn to the  right panel of Fig.\ref{Fig3}, where we have plotted the entanglement gap or Schmidt gap $\Delta$ for $J_y/J_x=0.3$. We find
in the gapped region that the entanglement entropy is vanishingly small but in the gapless regime the entanglement entropy
fluctuates as a function of $J_z$. However given the small value of $\Delta$, we can assume that it is essentially zero. We have also
looked at the zero energy eigenmode and found that they are essentially confined at the edge for small and large $J_z$, with small extensions to the bulk; 
\textit{this is very distinct from the case of the half-region where no gapless edge states were found in the small $J_z$ gapped phase}.
$\Delta =0 $ is a confirmation of  Refs. \onlinecite{haldane-2008,lukasz-2010} which posit that for a system with a gapped bulk,
the Schmidt gap depends on the zero energy edge modes. 
Moreover, analogous to the half-region of the previous section, although the system is gapless in the bulk for the shaded region, we 
can attribute the gapless edge modes as the cause of vanishingly small $\Delta$. \\
Now let us consider the Schmidt gap for $J_x=J_y$, which is shown in the inset of the right panel of Fig.\ref{Fig3}. 
We find a strong dependence of $\Delta$ on the subsystem geometry.
But they are more or less finite in the small $J_z$ limit where the system is gapless; $\Delta$ gradually becomes smaller as
$J_z$ approaches deep within the gapless phase, and remains zero thereafter. We have checked the lowest
eigenfunctions and found that although it is primarily localized at the edge it has considerable extension in the bulk
as well; in such a situation the precise relation between the edge mode and the entanglement gap $\Delta$ becomes complicated and is unclear to us.

\section{Analytical insights}
\label{section-4}
In this section we provide an analytical insight to corroborate our numerical findings in Secs. \ref{section-2} and \ref{section-3}.
It may be noted here that the results presented in the previous sections are exact
since in the derivation of Eqs. (\ref{final-ent}) and (\ref{final-gap}), no approximations have been used 
\cite{peschel-2003}. However, because of the complexity of obtaining an analytic expression for the correlation matrix and its spectral properties, one has to 
resort to numerical diagonalization of the same.

Using perturbative approximations in certain regimes of $J_z$ one may obtain some analytical insights into the nature of the entanglement entropy and gap,  
and may thus better understand the numerical results. \\
For completeness we first enlist our key results so far. 
Firstly, in the case of the half-region, we have found that the 
entanglement entropy increases parabolically with $J_z$ in the gapped phase (where $J_z$ is small i.e in 
the weakly coupled chain limit) and becomes oscillatory in the gapless regime.  
The entanglement gap is finite in the weakly coupled chain limit where the system 
is gapped. It goes to zero once the system enters the gapless regime and remains gapless in the large $J_z$ gapped 
phase, which we attributed to the presence of zero energy Majorana edge states. 
Secondly, we found that for $J_x=J_y$, there exists a critical value $J_z=J_c$ for the square block entanglement entropy, upto 
which it decreases and then either increases or remains constant. Thirdly, for the square block we found that 
the entanglement gap shows small fluctuations for small $J_z$. 
We begin our discussion in the small $J_z$ limit i.e weakly coupled chain limit in Sec. \ref{weak-lim} 
where we qualitatively explain the behaviour of the entanglement entropy and gap for the half-region. The 
behaviour of the entanglement entropy and gap in the large $J_z$ limit will be explained in Sec. \ref{large-lim}.

\subsection{Weakly coupled chain limit: small $J_z$}
\label{weak-lim}
The weakly coupled chain limit can be addressed by perturbation theory where the unperturbed
Hamiltonian consists of one-dimensional chains. The complete two-dimensional system may be considered as a composition
of many one-dimensional chains coupled by a small $J_z$.  Here we consider $N_y$ number of coupled chains with periodic boundary
so that the $N_y^{\textrm{th}}$  chain is connected to the first chain. The complete Hamiltonian for the two-dimensional 
lattice is written as
\begin{eqnarray}
&&\mathcal{H}= \mathcal{H}_0 + \mathcal{H}^{\prime}, \\
&&\mathcal{H}_0= \sum^{N_y}_{m} \mathcal{H}^m_0, ~~~~\mathcal{H}^{\prime}= \sum_m \mathcal{H}^{\prime}_{m,m+1}, \\
&&\mathcal{H}^m_0= \sum_{n} \Big( i J_x c^m_{n ,a} c^m_{n ,b}  + i J_y c^{m}_{n ,b} c^{m}_{n +1,a} \Big), \\
&&\mathcal{H}^{\prime}_{m,m+1}= \sum_{n}  i J_z c^{m}_{n ,a} c^{m-1}_{n ,b}.
\end{eqnarray}
Here  $\mathcal{H}^m_0$ is the unperturbed Hamiltonian for the $m^{\textrm{th}}$ one-dimensional chain and $\mathcal{H}^{\prime}$ 
denotes the interchain coupling and forms the perturbation to the unperturbed Hamiltonian $\mathcal{H}_0$. 
$\mathcal{H}$ can be diagonalised using a Fourier transform, $c^m_{n , \gamma} = \sum_{\vec{k}}\textrm{e}^{i\vec{k}.\vec{r}_{i, \gamma}}c^m_{n ,\gamma}$ 
for the $m^{\textrm{th}}$ chain where $\gamma = a, b$. 
Notice that for $c^m_{n }$ to be Majorana fermions, we must 
have $c^m_{-k,\gamma}= c^{m\dagger}_{k,\gamma}$ which implies that 
in momentum space only the fermion operators belonging to first half of Brillouin zone are independent i.e $\vec{k}-$ summation is over $(0,\pi)$. Using this definition 
and after subsequent diagonalization we can write, 
\begin{eqnarray}
\mathcal{H}^m_0 = \sum |\epsilon_k| \Big( \alpha^{m \dagger}_k \alpha^m_k - \beta^{m \dagger}_k \beta^m_k \Big),
\end{eqnarray}
with $\epsilon_k = |J_x + J_y e^{i k} |$. The new fermionic modes are defined as,
\begin{eqnarray}
\label{eq: substitution}
\left( \begin{array}{c}
c^m_{k,a} \\
c^m_{k,b} \end{array} \right)= \frac{1}{\sqrt{2}} \left( \begin{array}{cc}
i e^{i \theta_k} & i e^{i \theta_k} \\
1 & -1\end{array}  \right)
 \left( \begin{array}{c}
\alpha^m_k \\
\beta^m_k \end{array} \right),
\end{eqnarray}
where $ \theta_k =  \tan^{-1}\left( \frac{J_y\sin{k}}{J_x + J_y\cos{k}} \right)$. Using \eqref{eq: substitution}, the interchain 
perturbation  governed by $J_z$  can be written as,
\begin{eqnarray}
 \label{eq: perturbation}
\mathcal{H}^{m,m+1}_p &=&  \sum_{k \geq 0} \frac{J_z e^{i \theta_k}}{4} \Big( \alpha^{m \dagger}_{k} \alpha^{m-1}_{k} - \alpha^{m \dagger}_{k} \beta^{m-1}_{k} \nonumber \\
&& + \beta^{m \dagger}_{k} \alpha^{m-1}_{k} - \beta^{m\dagger}_{k} \beta^{m-1}_{k} \Big) + {\rm h.c},
 \end{eqnarray}
where the summation is over all the $m$ chains. 

Ground state of the system can be written as the product of the  individual ground states of each chain. Let $| g,m\rangle$
denote the ground state of $m^{th}$ chain and $| \mathcal{G} \rangle$ denote the  ground state for the complete
system, then we can write,
\begin{eqnarray}
| \mathcal{G} \rangle = \prod^{N_y}_{m=1} | g,m\rangle, ~~~~| g,m\rangle= \prod_k \beta^{m \dagger}_k |0 \rangle .
\end{eqnarray}

Our next task is to find the perturbed ground state when we take into account Eq.(\ref{eq: perturbation}) as the perturbation. 
For simplicity, we limit ourselves up to to second order in $J_z$ for the corrections in the entanglement entropy and gap 
and neglect subsequent higher order corrections.
For this purpose it is sufficient to limit up to the first order corrections of the ground state which is given as,

\begin{eqnarray}
\label{1stcorr}
|\mathcal{G}_1 \rangle &&= |\mathcal{G} \rangle \Big( 1- N_y\sum_k  \frac{J^2_z}{64} \frac{1}{\epsilon^2_k} \Big) \nonumber \\
-\sum_m  &&\frac{(-1)^{N_y-m}J_z}{8 \epsilon_k}  \Big(e^{-i \theta_k} | 0,0;m-1 \rangle  | 1,1;m \rangle |\mathcal{G}:m,m-1 \rangle \nonumber \\
&& - e^{i \theta_k} | 0,0;m+1 \rangle  | 1,1;m \rangle |\mathcal{G}:m+1,m \rangle \Big).
\end{eqnarray}
 
Where, $|\mathcal{G}:a,b,c,...\rangle = \prod_{m \neq a,b,c,..} | g,m \rangle$ and   $| p,q;m \rangle = (\alpha^{\dagger}_m)^p (\beta^{\dagger}_m)^q |0 \rangle$, 
where $p,~ q$ can only take values $0$ or $1$. The reduced density matrix of the half-region can be easily calculated by considering the first order correction to the ground state wavefunction itself. 
This contains two diagonal terms: the first comes from the first term of Eq.(\ref{1stcorr}) and is proportional to unity. 
The second term of Eq.(\ref{1stcorr}) yields another diagonal term proportional to $J^2_z$. 
The off-diagonal terms are of two types, terms which are proportional to $J_z$ and is obtained from the product of first and third terms of Eq.(\ref{1stcorr}). 
These off-diagonal terms can be treated as perturbation to the first diagonal terms of the reduced density matrix which is proportional to unity. 
While the other off-diagonal terms coming from the second term yields a correction $\sim J^4_z$. Thus considering only the terms upto $J^2_z$, we get the following expressions for 
the largest and second largest eigenvalues of the reduced density matrix,

\begin{eqnarray}
\label{eq:red-eig-value}
&&\lambda_{1}= \lambda_0 + \sum_k\frac{(N^{\prime}-1)J^2_z}{32 \epsilon^2_k} \Big(\lambda_0- \frac{J^2_z}{64 \epsilon^2_k} \Big)^{-1} ,\\
&&\lambda_2= \Big( \frac{J^2_z}{64 \epsilon^2_{k}} \Big)_{\rm min}=\frac{J^2_z}{64 (J_x-J_y)^2} ,
\end{eqnarray}
where $\lambda_0=1- N_y\sum_k \frac{J^2_z}{32 \epsilon^2_k}$. 
In the above $N^{\prime}$ 
is the number of chains in the subsystem. From the expression of Eq.(\ref{eq:red-eig-value}), we find that for small values of $J_z$, the 
entanglement entropy has a parabolic dependence on $J_z$ as is found in the upper left panel of Fig.\ref{Fig2}.

\subsection{Dimer limit: strong $J_z$}
\label{large-lim}
\label{half-chain2}
\begin{figure}
\includegraphics[width=0.5\textwidth]{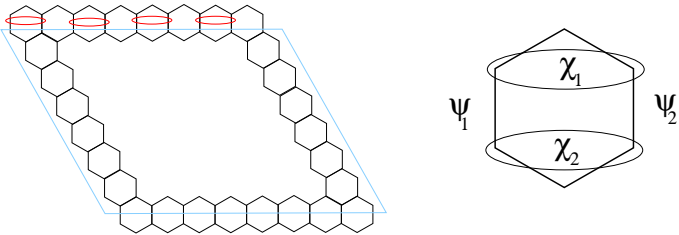}
\caption{(Colour online) 
In the large $J_z$ limit we associate each $z$-bond with a $\psi$ fermion in order to compute the reduced density matrix.
The boundary is composed of the upper, lower, and the two side boundaries. 
For the upper and lower boundaries a $\psi$ fermion is shared between the system and the subsystem. For this reason we regrouped the two adjacent $\psi$ fermions to define 
two $\chi$ fermions (as shown at the right side of the figure) such that $\chi_1$ belongs to the system and $\chi_2$ belongs to the subsystem.}
\label{majo-group}
\end{figure}
 
In this section we present the perturbative results
for entanglement entropy and gap in the large $J_z$ limit for the half-region. In the limit $J_z \rightarrow \infty$  the Hamiltonian 
consists of isolated $z$-bonds only and the Hamiltonian for each z-bond is $J_z i c_{k,1} c_{k,2}$ where $k$ denotes a particular $z$-bond 
and `1' and `2' refer to the sites inside and outside of the subsystem. It is straightforward to see that integrating out one Majorana fermion 
yields a contribution $ {\rm log} 2$  to the entropy. Thus if there are in total $N_z$ number of $J_z$ bonds shared between the subsystem 
and the system, we obtain the limiting values of the entanglement entropy as $N_z {\rm log} 2$. The perturbation here amounts to switching on the hopping of Majorana fermions 
between nearest-neighbour dimers. The Hamiltonian can then be written as
\begin{equation}
H= \sum_{n}  J_z i c_{n ,1} c_{n ,2} +  \sum_{n } J_{\alpha} i c_{n ,1} c_{n +\delta_\alpha,2},~\alpha=x,y .
\end{equation} 

In the above expression $n +\delta_{\alpha}$ refers to the  nearest-neighbour dimer connected by an $\alpha$-bond with the $n^{\textrm{th}}$
dimer. 
We may now define a $\psi$ fermionic basis using the substitution $c_{n ,1}= \psi_{n}  + \psi^{\dagger}_{n} ,~~~~ c_{n ,2}= -i (\psi_{n}  - \psi^{\dagger}_{n}  )$, 
such that $\psi_n |0\rangle_n = 0$, with $|0\rangle_n$ being the vacuum state of the dimer at site `$n$'; 
the ground state of the full system is then obtained as $|\mathcal{G}_0 \rangle=|\mathcal{O}\rangle$ where $|\mathcal{O}\rangle= \prod_i|0\rangle_i$.  
To calculate the reduced density matrix, as in the weak $J_z$ limit,
we begin with the  perturbed ground state. The perturbed ground state to second order in $J = J_{\alpha},~~\alpha=x,y$, is then given by
\begin{equation}
|\mathcal{G}_1 \rangle= \left(1- \tilde{N} \frac{J^2}{8J^2_z}\right)|\mathcal{O}\rangle + \sum_{<i,i+\delta_{\alpha}>} \frac{J_{\alpha}}{2 J_z}|1_{n} , 1_{{n} +\delta_{\alpha}} \rangle,  
\end{equation}
where $|1_{n}, 1_m \rangle= |1\rangle_{n}  \times |1\rangle_{m} $ and denotes the filled states at the dimer `$n $' and `$m $' and $\tilde{N}$ is the total number of $z$-bonds in the system.
We retain the expansion of the ground state upto first order as we intend to find the reduced density matrix upto second
order in $J$.  We see that the two Majorana fermions
may be grouped together to define a single complex fermion at a given $z$-bond. While calculating the reduced density matrix one
needs to integrate out the Majorana fermion outside the subsystem. To this end we introduce a pair of $\chi$ fermions $(\chi_1, \chi_2)$ out of the
two adjacent $\psi$ fermions as shown in the Fig.\ref{majo-group} and then calculate the reduced density matrix of
an extended subsystem which includes both the sites of the boundary $z$-bonds. 
The reduced density matrix for the half-region can then be obtained easily by taking 
a trace of the $\chi_1$ fermions. The detailed mapping between the Fock space of $\psi_{1,2}$ and $\chi_{1,2}$ fermions is as follows: 
\begin{eqnarray}
\label{eq: mapping}
 |0,0\rangle_{\psi_j}&=& \frac{1}{\sqrt{2}} (|1,0\rangle_{\chi_j} + i|0,1\rangle_{\chi_j} ), \nonumber \\
 |1,0\rangle_{\psi_j}&=& \frac{1}{\sqrt{2}} (|0,0\rangle_{\chi_j} - i|1,1\rangle_{\chi_j} ), \nonumber \\
 |0,1\rangle_{\psi_j}&=&- \frac{1}{\sqrt{2}} (i |0,0\rangle_{\chi_j} - i|1,1\rangle_{\chi_j} ),\nonumber \\
 |1,1\rangle_{\psi_j}&=& \frac{1}{\sqrt{2}} (|0,1\rangle_{\chi_j} + i|1,0\rangle_{\chi_j} ),
\end{eqnarray}

where $j=1,2$. We can now write down the reduced density matrix of the extended subsystem in the $\psi$ basis as follows 
\begin{eqnarray}
\rho_E&&= \left(1- N \frac{J^2}{4J^2_z}\right) |\mathcal{O}\rangle \langle \mathcal{O}| + \sum_{i} \Big( \frac{J^2}{4 J^2_z} |\tilde{\mathcal{O}},1_i\rangle \langle \tilde{\mathcal{O}},1_i|  +\nonumber \\
&& \sum_{i,j} \left[ \frac{J^2}{4 J^2_z} |\tilde{\mathcal{O}},1_i, 1_j\rangle \langle \tilde{\mathcal{O}},1_i,1_j| + \frac{J}{2 J_z} |\mathcal{O}\rangle \langle \tilde{\mathcal{O}},1_i,1_j| \right] \Big). \nonumber \\
\label{red-extnd}
\end{eqnarray}
In the above expression $|\mathcal{O}\rangle $ denotes the vacuum of the  extended subsystem only 
and $|\tilde{\mathcal{O}},1_i,1_j...\rangle $ denotes a state with filled dimer on sites $i,~j,$ etc. 
In the second term, the index `$i$' refers only to the boundary $z$-bonds. 
In third and fourth terms `$i$' and `$j$' denotes nearest-neighbours.  Now employing the transformation in Eq.(\ref{eq: mapping}) between  $\psi$ and $\chi$
basis, we immediately see that first term of Eq.(\ref{red-extnd}) yields $N_z$ number of degenerate eigenvalues 
$\frac{1}{2} (1- \tilde{N} \frac{J^2}{4J^2_z})$ where $N_z$ is the number of boundary z-bonds. This is a vital point and corroborates the vanishing Schmidt gap observed in the dimer 
limit. This result, taken together with the analytical computations of Ref. \onlinecite{diptiman} where zig-zag Majorana edge states are unveiled in the large-$J_z$ limit, provides 
an analytical confirmation of the bulk-edge correspondence \cite{haldane-2008, lukasz-2010}.\\
Furthermore these eigenvalues are 
obtained from the diagonal part of the reduced density matrix and the fourth term in Eq.(\ref{red-extnd}) is the perturbation acting on these 
terms. The second and third terms of Eq.(\ref{red-extnd}) yields the second largest eigenvalues and they are 
also diagonal. After some straightforward algebra we find the largest and the second largest eigenvalues to be given by
\begin{eqnarray}
\label{eq:half-eig-value}
&&\lambda_{1}= \lambda_0 + \frac{\tilde{N} J^2}{64 J^2_z} \Big(\lambda_0- \frac{J^2}{64 J^2_z} \Big)^{-1} ,\\
&&\lambda_2=  \frac{J^2}{8 J^2_z},
\end{eqnarray}
where $\lambda_0=1- {\tilde N} \frac{J^2}{2 J^2_z}$.
\section{Conclusions}
\label{section-5}

We have presented an extensive study of the entanglement entropy and Schmidt gap for the vortex-free ground state of the Kitaev model,
and showed its variation along specific contours in the phase space of parameters. The separation of the contributions of the 
entanglement entropy into a free (Majorana) fermionic part and a gauge field part allowed us to treat large systems. We have considered
mainly two  specific geometries \textit{viz.} 
a square/rectangular block and the half-region, the latter being defined as one half of the torus. For both geometries of the subsystem, the free fermionic 
entanglement entropy is found to capture the presence of phase transitions in the Kitaev model. \\
For the half-region the entanglement entropy was found to be an oscillating function
in the gapless phase due to the long range correlation that exists in the gapless phase and the presence of gapless $\vec{k}$-modes. 
In the gapped phases such oscillatory behaviour is absent: we find the entanglement entropy to monotonically vary with the coupling $J_z$, albeit decreasing in the Toric code limit and increasing in the weakly coupled chain limit with 
increasing $J_z$.
The entanglement gap, on the other hand, is finite in the weakly coupled gapped phase (i.e when $J_z$ is small and $J_x \neq J_y$ ) and drops to zero as soon as the system enters the gapless phase 
satisfying $J_x \leq J_y + J_z$ and its cyclic combinations.
The entanglement gap remains zero even when the interchain coupling $J_z$ is increased further 
and the system enters the Toric code limit ($J_{z} \geq J_{x} + J_y$) which is also gapped. Thus we have shown that, although both the 
gapped regions are of the same topological character (as explained  earlier in the text), the entanglement
 gap is finite in one phase and zero in the other. We have demonstrated how this may be attributed to the presence or absence of zero energy edge modes in the system, 
 with the latter's connection with the entanglement gap being further confirmation of the bulk-edge correspondence.\\
For the square/rectangular block the entanglement entropy was seen to exhibit nonmonotonic 
behaviour as a function of interchain coupling $J_z$ within in the gapless phase particularly for $J_x=J_y$; 
generically reaching a point of minimum entanglement within the gapless phase before the entanglement starts increasing 
again with the coupling. For $J_x \neq J_y$ the qualitative behaviour of the entanglement entropy in the gapless phase depends on the ratio of $J_y/J_x$ and 
$J_z/J_x$ . We explained that this is due
to the competition between correlation functions in different kind of bonds that are shared between the subsystem
and rest of the system. In the gapless phase of the weakly coupled chain limit, the entanglement gap is a mildly fluctuating function of the interchain
coupling parameter $J_z$ due, possibly, to the finite extent of edge modes into the bulk.  \\
We have corroborated our numerical findings with perturbative 
analytical calculations in the appropriate regimes, in particular corroborating the bulk-edge correspondence for the half-region. 
A more detailed analytical study of the entanglement gap for the two regions deserves further attention, especially close to the transitions, which we leave for future work.

One of the authors (VKV) thanks F. Franchini and T. Grover for discussions.

\end{document}